\documentclass[11pt]{llncs}
\usepackage{amssymb,bbm}
\usepackage{tikz}
\usetikzlibrary{matrix,arrows}
\usepackage{wrapfig}
\usepackage{listings}
\usepackage{stmaryrd,mathtools}
\usepackage{url}

\title{Theory Presentation Combinators\thanks{This research was
    supported by NSERC.}}

\author{Jacques Carette \and Russell O'Connor
    \thanks{\texttt{carette@mcmaster.ca},\texttt{roconn@mcmaster.ca}.}}

\institute{%
Department of Computing and Software\\
McMaster University\\
Hamilton, Ontario, Canada
}

\newcommand{\thy}[1]{\texttt{#1}}
\newcommand{\C}{\mathbbm{C}}
\newcommand{\B}{\mathbbm{B}}
\newcommand{\E}{\mathbbm{E}}
\newcommand{\V}{\mathbbm{V}}
\newcommand{\N}{\mathbbm{N}}
\newcommand{\EmpThy}{\left\langle\,\right\rangle}

\newcommand{\sem}[2]{\ensuremath{\llbracket{#2}\rrbracket_{#1}}}
\newcommand{\semC}[1]{\sem{\B}{#1}}
\newcommand{\semE}[1]{\sem{\E}{#1}}
\newcommand{\semP}[1]{\sem{\pi}{#1}}
\newcommand{\partialf}{\rightharpoonup}

\newcommand{\ctx}[4]{\ensuremath{\left\langle #1:#2 \right\rangle_{#3}^{#4}}}
\newcommand{\ren}[4]{\ensuremath{\left[ #1\mapsto #2 \right]_{#3}^{#4}}}
\newcommand{\ctxcat}{\fatsemi}
\newcommand{\rencat}{\Join}

\newcommand{\tmop}[1]{\ensuremath{\operatorname{#1}}}
\newcommand{\tmtexttt}[1]{{\ttfamily{#1}}}
\newcommand{\assign}{:=}
\newcommand{\tmdfn}[1]{\textbf{#1}}

\begin{document}
\maketitle

\lstdefinelanguage{mathscheme}
    {morekeywords={Theory,combine,extended,by,type,over,axiom,
        implies,not,forall,and,not,or,Inductive,case,of,instance,via,
        defined-in},
    basicstyle={\small},
    keywordstyle=\bfseries}

\lstset{language=mathscheme}

\begin{abstract}
We motivate and give semantics to \emph{theory presentation combinators}
as the foundational building blocks for a scalable library of theories.
The key observation is that the \emph{category of contexts} and
fibered categories are the ideal theoretical tools for this purpose. 
\end{abstract}

\section{Introduction}
A mechanized mathematics system, to be useful, must possess a large
library of mathematical knowledge, on top of sound foundations.  While
sound foundations contain many interesting intellectual challenges, building
a large library seems a daunting task because of its
sheer volume.  However, as has been 
well-documented~\cite{MathSchemeExper,CaretteKiselyov11,DuplicationMizar}, 
there is a tremendous amount of redundancy in existing libraries.

Our aim is to build tools that allow library developers to take
advantage of all the commonalities in mathematics so as to build 
a large, rich library for end-users, whilst expending much less actual
development effort.  In other words, we continue with our approach of
developing \emph{High Level Theories}~\cite{CaretteFarmer08} through building
a network of theories, by putting our previous
experiments~\cite{MathSchemeExper} on a sound theoretical basis.

\subsection{The Problem}

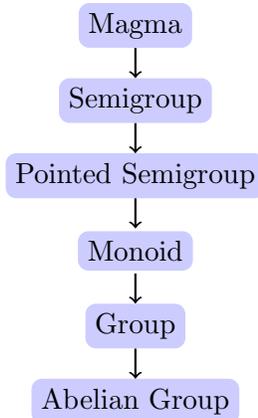
\begin{wrapfigure}[14]{L}{0pt}
\begin{tikzpicture}
\coordinate (M1) at (3, 5);
\coordinate (M2) at (3, 4);
\coordinate (M3) at (3, 3);
\coordinate (M4) at (3, 2);
\coordinate (M5) at (3, 1);
\coordinate (M6) at (3, 0);
\node[fill=blue!20,rounded corners] at (M1) {Magma};
\node[fill=blue!20,rounded corners] at (M2) {Semigroup};
\node[fill=blue!20,rounded corners] at (M3) {Pointed Semigroup};
\node[fill=blue!20,rounded corners] at (M4) {Monoid};
\node[fill=blue!20,rounded corners] at (M5) {Group};
\node[fill=blue!20,rounded corners] at (M6) {Abelian Group};
\draw [->,thick] (3,4.7) -- (3,4.3);
\draw [->,thick] (3,3.7) -- (3,3.3);
\draw [->,thick] (3,2.7) -- (3,2.3);
\draw [->,thick] (3,1.7) -- (3,1.3);
\draw [->,thick] (3,0.7) -- (3,0.3);
\end{tikzpicture}
\caption{Theories}\label{fig:thystruct1}
\end{wrapfigure}

The problem which motivates this research is fairly simple: give 
developers of mathematical libraries the foundational tools they need to
take advantage of the inherent structure of mathematical theories, as 
first class mathematical objects in their own right.  
Figure~\ref{fig:thystruct1} shows the type of structure we are talking about:
The presentation of the theory \thy{Semigroup} strictly contains that of
the theory \thy{Magma}, and this information should not be duplicated.
A further requirement is that we need to be able to selectively hide (and
reveal) this structure from end-users.

The motivation for these tools should be obvious, but let us nevertheless
spell it out: we simply cannot afford to spend the human resources necessary
(one estimate was $140$ person-years~\cite{Wiedijk};~\cite{AspertiCoen} explore
this topic in much greater depth) to develop yet another
mathematical library.  In fact, as we \emph{now} know that there is a lot of 
\emph{structured} redundancy in such libraries, it would be downright foolish
to not take full advantage of that.  As a minor benefit, it can also help
reduce errors in axiomatizations.

The motivation for being able to selectively hide or reveal some of this
structure is less straightforward.  It stems from our
observation~\cite{CaretteFarmer08} that \textit{in practice}, when
mathematicians are \emph{using} theories rather than developing news ones, they
tend to work in a rather ``flat'' name space.  An analogy: someone working in
Group Theory will unconsciously assume the availability of all concepts from a
standard textbook, with their ``usual'' names and meanings.  As
their goal is to get some work done, whatever structure system builders have
decided to use to construct their system should not leak into the application
domain.  They may not be aware of the existence of pointed semigroups, nor
should that awareness be forced upon them.  Some application domains
rely on the ``structure of theories'', so we can allow those users to see it.

\subsection{Contributions}
To be explicit, our contributions include:
\begin{itemize}
\item A variant of the \emph{category of contexts}, over a dependently-typed
type theory as the semantics for theory presentations.
\item A simple term language for building theories, using ``classical''
nomenclature, even though our foundations are unabashedly categorical.
\item Using ``tiny theories'' to allow for maximal
reuse and modularity.
\item Taking names seriously, since these are meant for human consumption.
Moreover, we further emphasize that theory presentations are purely
syntactic objects, which are meant to \emph{denote} a semantic object.
\item Treating arrows seriously: while this is obvious from
a categorical standpoint, it is nevertheless novel in this application.
\item Giving multiple (compatible) semantics to our language, which
better capture the complete knowledge context of the terms.
\end{itemize}

\subsection{Plan of paper}

We motivate our work with concrete examples in
section~\ref{sec:tpc1}.  The theoretical foundations of our work, the
fibered category of contexts, is presented in full detail in 
section~\ref{sec:category}.  This allow us in section~\ref{sec:tpc2} to
formalize the language of our motivation section, syntactically and
semantically.  We close with some discussion, related work and conclusions
in sections~\ref{sec:discussion}--\ref{sec:conclusion}.

\section{Motivation for Theory Presentation Combinators}\label{sec:tpc1}
Let us compare the presentation of two simple theories:
\begin{lstlisting}
Monoid := Theory { 
  U:type;  *:(U,U) -> U;  e:U;
  axiom rightIdentity_*_e: forall x:U. x*e = x;
  axiom leftIdentity_*_e: forall x:U. e*x = x;
  axiom associative_*: forall x,y,z:U. (x*y)*z = x*(y*z)}

CommutativeMonoid := Theory { 
  U:type;  *:(U,U) -> U;  e:U;
  axiom rightIdentity_*_e: forall x:U. x*e = x;
  axiom leftIdentity_*_e: forall x:U. e*x = x;
  axiom associative_*: forall x,y,z:U. (x*y)*z = x*(y*z);
  axiom commutative_*: forall x,y:U. x*y = y*x}
\end{lstlisting}

\noindent  They are identical, save for the \thy{commutative\_*}
axiom, as expected.  Given \thy{Monoid}, it would be much more 
economical to define
\begin{lstlisting}
CommutativeMonoid := Monoid extended by {
  axiom commutative_*: forall x,y:U. x*y = y*x}
\end{lstlisting}

\noindent and ``expand'' this definition, if necessary.  Of course, given
\thy{Group}, we would similarly find ourselves writing
\begin{lstlisting}
CommutativeGroup := Group extended by {
  axiom commutative_*: forall x,y:U. x*y = y*x}
\end{lstlisting}
\noindent which is also wasteful, as well as dangerous: is this ``the same''
axiom as before, or a different one?  There is no real way to tell.  It is
natural to further extend our language with a facility that expresses 
this sharing.  Taking a cue from previous work, we might want to say
\begin{lstlisting}
CommutativeGroup := combine CommutativeMonoid, Group over Monoid
\end{lstlisting}
\noindent  Informally, this can be read as saying that \thy{Group} and
\thy{CommutativeMonoid} are both ``extensions'' of \thy{Monoid}, and
\thy{CommutativeGroup} is formed by the union (amalgamated sum) of those
extensions.  Another frequent feature is \emph{renaming}: an
\thy{AbelianGroup}, while isomorphic to a \thy{CommutativeGroup}, is 
usually presented additively.  We could express this as
\begin{lstlisting}
AbelianGroup := CommutativeGroup[ * |-> +, e |-> 0 ]
\end{lstlisting}

Unfortunately, while this ``works'' to build a sizeable library (say of the
order of $500$ concepts) in a very economical way, it is quite brittle.  Let
us examine the reasons.  It should be clear that by \thy{combine}, we
really mean \emph{pushout}\footnote{Following Burstall and 
Goguen~\cite{PuttingTheoriesTogether} and
Smith~\cite{Smith93,Smith99} and many others since.}.  But a pushout is 
a $5$-ary operation on $3$ objects and $2$ arrows; our syntax gives the $3$
objects and leaves the arrows implicit.  This is a very serious mistake:
these arrows are (in general) not easy to infer, especially in the presence of
renaming.  For example, there are two distinct arrows from \thy{Monoid} to 
\thy{Ring}, with neither arrow being ``better'' than the other.
Furthermore, we know that pushouts can also be regarded as a
$2$-ary operation on arrows.  In other words, even though our goal is to
produce \emph{theory presentations}, our decision to use pushouts%
\footnote{which will in fact become pullbacks} as a
fundamental building block gives us no choice but to 
\textbf{take arrows seriously}.

So our task is now to find a category with ``theory presentations'' as objects,
and with arrows which somehow express the notions of extending, combining
and renaming as defined above.  But before we explore that in depth, let us
further examine our operations.  First, there is nothing specific to
\thy{CommutativeGroup} in the renaming $* \mapsto +, e \mapsto 0$, this
can be applied to any theory where the pairs $(*,+)$ and $(e,0)$ have 
compatible signatures (including being undefined).  Similarly, 
\thy{extend} really defines a ``construction'' which can be applied 
whenever all the symbols used in the extension are defined.  In other words,
a reasonable semantics should associate a whole class of arrows%
\footnote{We are being deliberately vague here, Section~\ref{sec:category} will
make this precise.} to these operations.

But there is one more aspect to consider: in all our examples above,
we have used short, meaningful names.  While great for humans, they are
in part at fault in the failure of being able to infer arrows.  If, like in
MMT~\cite{MMT}, we used long names, might we be able to 
build a robust system?  Maybe so, but it would immediately fall afoul of
our second requirement: irrelevant information such as choices made by
developers regarding the order in which to build theories, would leak into
the long names, and thus be seen by users.  Furthermore, when there is
ambiguity, a long name system can indeed resolve that ambiguity,
but at too high a cost to humans in absurdly long names for certain concepts.

In other words, to be able to maintain human-readable names for all
concepts, we will put the burden on the \emph{library developers} to come up
with a reasonable naming scheme, rather than to push that issue onto end users.
Another way to see this is that symbol choice carries a lot of intentional, 
as well as contextual, information which is commonly used in mathematical
practice.  Thus, to avoid leaking irrelevant information and to maintain
intentional/contextual information, we will insist that on
\textbf{taking names seriously}.

\section{Category of contexts}\label{sec:category}

We observe that theories from the previous section can all be specified as
contexts of some dependent type theory.  The work in this paper is abstract
over the exact details of the dependent type theory,\footnote{In fact, we expect
this work to apply not only to dependent type theories, but to any classifying
category~\cite{JacobsCLTT}.} so we simply assume that some dependent type theory
is given. Following Cartmell~\cite{Cartmell}, we form the category of contexts
$\C$ of the given dependent type theory.
The objects of $\C$ are contexts $\Gamma$ that occur
in judgements like $\Gamma \vdash s : \sigma$ of the dependent type theory. A
context $\Gamma$ consists of a sequence of pairs of labels and types
 (or kinds or propositions),
\[ \Gamma \assign \left\langle x_0 : \sigma_0 ; \ldots ; x_{n - 1} : \sigma_{n
   - 1} \right\rangle, \]
such that for each $i < n$ the judgement 
\[\left\langle x_0 : \sigma_0 ; \ldots
; x_{i - 1} : \sigma_{i - 1} \right\rangle \vdash \sigma_i : \tmop{Type}\]
\noindent holds (resp. $: \tmop{Kind}$, or $: \tmop{Prop}$). Contexts of dependent
type theory can be used to define the types, operations, relations and axioms
of a theory.  We will use the
abbreviation $\ctx{x}{\sigma}{0}{n-1}$ for a context $\Gamma$, and $\ctxcat$
for concatenation of two such sequences.

\begin{example}
  \label{SemigroupExample}
  We can define the theory of semigroups via
  \[ \mathtt{Semigroup} \assign \left\langle \begin{array}{rcl}
       U & : & \tmop{Type}\\
       \left( * \right) & : & U \times U \rightarrow U\\
       \tmop{associative} & : & \forall x, y, z : U. \left( x * y
       \right) * z = x * \left( y * z \right)
     \end{array} \right\rangle \]
\noindent where we use Haskell-style notation where $\left( \Box \right)$
indicates (the name of) a binary function used infix in terms.
\end{example}

Normally contexts are considered up to $\alpha$-equivalence, that is, 
renaming or permuting the labels of a context makes no difference.
But since labels do make a difference, we will not do so.  However,
$\alpha$-equivalent \emph{terms} and \emph{types} continue to be considered
equivalent.

\begin{example}
  \label{AdditiveSemigroupExample}The signature for 
    \tmtexttt{AdditiveSemigroup} is given as the context
    \[ \left\langle \begin{array}{rcl}
       U & : & \tmop{Type}\\
       \left( + \right) & : & U \times U \rightarrow U\\
       \tmop{associative} & : & \forall x, y, z : U. \left( x + y \right) + z
       = x + \left( y + z \right)
     \end{array} \right\rangle  \]
  Traditionally \tmtexttt{Semigroup} and \tmtexttt{AdditiveSemigroup} would be
  considered the same context because they are $\alpha$-equivalent.
\end{example}

In the rest of this section, we will use the convention that
$\Gamma = \ctx{x}{\sigma}{0}{n-1}$ and $ \Delta = \ctx{y}{\tau}{0}{m-1}$.
Given two contexts $\Gamma$ and $\Delta$, a morphism $\Gamma \rightarrow
\Delta$ of $\C$ consists of an assignment $\left[ y_0 \mapsto t_0,
\ldots, y_m \mapsto t_{m - 1} \right]$, abbreviated as
$\ren{y}{t}{0}{m-1}$ where the $t_0, \ldots, t_{m - 1}$ are
terms such that
\[
  \Gamma \vdash t_0 : \tau_0\qquad \ldots 
  \qquad \Gamma \vdash t_{m - 1} : \tau_{m - 1}\ren{y}{t}{0}{m-2}
\]
all hold, where $\tau \ren{y}{t}{0}{i}$
denotes the type $\tau$ with the labels $y_0, \ldots, y_i$ substituted by the
corresponding terms of the assignment.  We will also use $\rencat$ to denote
concatenation of assignments, and $\ren{y_{f(j)}}{t_{g(j)}}{j=a}{b}$ for
the ``obvious'' generalized assignment.

Notice that an arrow from $\Gamma$ to $\Delta$ is an assignment from the labels
of $\Delta$ to terms in $\Gamma$.  This definition of an arrow may seem
backwards at first, but it is defined this way because arrows  transform
``models'' of theories of $\Gamma$ to ``models'' of theories of $\Delta$.
For example, every Abelian Semigroup is, or rather can be transformed
into, an Additive Semigroup by simply forgetting that the Semigroup is Abelian.
A later example~\ref{ex:AbelianSemigroupToAdditiveSemigroup} will give the 
explicit arrow from Abelian Semigroup to Additive Semigroup that captures this
transformation.

Let us fix $\V$ as the (countable) infinite set of labels used in
contexts.  If $\pi : \V \rightarrow \V$ is a permutation of
labels, then we can define an action of this permutation on terms, types and
contexts:
\[ \pi \cdot \ctx{x}{\sigma}{0}{n-1}
   \assign \left\langle \pi \left( x_0 \right) : \pi \cdot
   \sigma_0 ; \ldots ; \pi \left( x_{n - 1} \right) : \pi \cdot \sigma_{n - 1}
   \right\rangle  \equiv \ctx{\pi x}{\pi\cdot\sigma}{0}{n-1}\]
where $\pi \cdot \sigma_i$ is the action induced on the (dependent) type
$\sigma_i$ by renaming labels. The action of $\pi$ induces an endofunctor
$\left( \pi \cdot - \right) : \C \rightarrow \C$.
Furthermore, each permutation $\pi : \V \rightarrow \V$
induces a natural transformation in $I_{\pi} : \left( \pi \cdot - \right)
\Rightarrow \tmop{id}_{\C}$ where
\[ I_{\pi} \left( \Gamma \right) \assign \left[ x_0 \mapsto \pi \left( x_0
   \right), \ldots, x_{n - 1} \mapsto \pi \left( x_{n - 1} \right) \right] :
   \pi \cdot \Gamma \rightarrow \Gamma . \]
We call an assignment of the form $I_{\pi} \left( \Gamma \right)$ a
{\tmdfn{renaming}}. Because permutations are invertible, each renaming
$I_{\pi} \left( \Gamma \right) : \pi \cdot \Gamma \rightarrow \Gamma$ is an
isomorphism whose inverse is the renaming $I_{\pi^{- 1}} \left( \pi \cdot
\Gamma \right) : \Gamma \rightarrow \pi \cdot \Gamma$. From this we can see
that $\alpha$-equivalent contexts are isomorphic.

\begin{example}
  Let $\pi : \V \rightarrow \V$ be some permutation such
  that
    $\pi \left( U \right) = U$,\ 
    $\pi \left( \left( * \right) \right) = \left( + \right)$, and
    $\pi \left( \tmop{associative} \right) = \tmop{associative}$.
  By the definition of $I_{\pi} \left( \mathtt{Semigroup} \right) :
  \mathtt{AdditiveSemigroup} \rightarrow \mathtt{Semigroup}$, we
  have that
  \[ I_{\pi} \left( \mathtt{Semigroup} \right) \assign \left[ U \mapsto
     U ; \left( * \right) \mapsto \left( + \right) ; \tmop{associative}
     \mapsto \tmop{associative} \right] \]
  is a renaming isomorphism between the contexts in examples
  \ref{SemigroupExample} and \ref{AdditiveSemigroupExample}.
\end{example}

The category of nominal assignments, $\B$, a sub-category of $\C$ will 
be quite important for use.  For example, theorem~\ref{codIsFibration} will
show that $\B$ is the base category of a fibration.

\begin{definition}
  The category of nominal assignments, $\B$, has the same objects 
  as $\C$, but only those morphisms whose terms are labels.
\end{definition}

\noindent Thus a morphism in $\B$ is an
assignment of the form $\ren{y_i}{x_{a\left(i\right)}}{i=0}{m-1}$
such that the judgements
\[
  \Gamma \vdash x_{a \left( 0 \right)} : \tau_0\qquad \ldots
  \qquad \Gamma \vdash x_{a \left( m - 1 \right)} : \tau_{m - 1} 
    \ren{y_i}{x_{a\left(i\right)}}{i=0}{m-2}
\]
all hold. 
\begin{definition}
  We define $\Gamma$ to be a {\tmdfn{sub-context}} of $\Gamma^+$ if every
  element $x : \tau$ of $\Gamma$ occurs in $\Gamma^+$.
\end{definition}

\begin{definition}
  We call an assignment $A : \Gamma \rightarrow \Delta$ a diagonal
  assignment if $A$ is of the form $\ren{y}{y}{0}{n-1}$ (where 
$\Delta = \ctx{y}{\tau}{0}{n-1}$),  denoted 
  by $\delta_{\Delta} : \Gamma \rightarrow \Delta$.
\end{definition}

\begin{definition}
  An assignment $A : \Gamma^+ \rightarrow \Gamma$ is an \emph{extension} when 
    $\Gamma$ is a sub-context of $\Gamma^+$, and A is the diagonal assignment.
\end{definition}

Notice that an extension points from the extended context to the sub-context.
This is the reverse from what Burstall and 
Goguen~\cite{PuttingTheoriesTogether} use (and most of the algebraic 
specification community followed their lead). Our direction
is inherited from $\C$, the category of contexts, which
is later required by theorem~\ref{codIsFibration} to satisfy the technical
definition of a fibration.

\begin{example}\label{ex:AbelianSemigroupToAdditiveSemigroup}
  Consider the theory $\mathtt{AbelianSemigroup}$ given as
  \[ \left\langle
     \begin{array}{rcl}
       U & : & U : \tmop{Type}\\
       \left( + \right) & : & U \times U \rightarrow U\\
       \tmop{associative} & : & \forall x, y, z : U. \left( x + y \right) + z
       = x + \left( y + z \right)\\
       \tmop{commutative} & : & \forall x, y : U. \left( x + y \right) =
       \left( y + x \right)
     \end{array} \right\rangle \]
  Then 
  $\delta_{\text{\tmtexttt{AdditiveSemigroup}}} :
     \mathtt{AbelianSemigroup} \rightarrow
     \mathtt{AdditiveSemigroup}$
  is an extension.
\end{example}

\begin{example}
  Consider the following two distinct contexts ($\mathtt{C}_1, \mathtt{C}_2$) 
  for the theory of left unital Magmas with
  the order of their operators swapped:
  \[
    \left\langle \begin{array}{rcl}
       U & : & \tmop{Type}\\
       e & : & U\\
       \left( * \right) & : & U \times U \rightarrow U\\
       \tmop{leftIdentity} & : & \forall x : U.e * x = x\\
     \end{array} \right\rangle \qquad
     \left\langle \begin{array}{rcl}
       U & : & \tmop{Type}\\
       \left( * \right) & : & U \times U \rightarrow U\\
       e & : & U\\
       \tmop{leftIdentity} & : & \forall x : U.e * x = x\\
     \end{array} \right\rangle \]
  The diagonal assignment $\delta_{C_1} : \mathtt{C}_2
  \rightarrow \mathtt{C}_1$ is an extension (as is $\delta_{C_2} :
    \mathtt{C}_1\rightarrow \mathtt{C}_2$).
\end{example}

Notice that, for any given contexts $\Gamma^+$ and $\Gamma$, there exists an
extension $\Gamma^+ \rightarrow \Gamma$ if and only if $\Gamma$ is a
sub-context of $\Gamma^+$. If $\Gamma$ is a sub-context of $\Gamma^+$ then the
diagonal assignment $\delta_{\Gamma} : \Gamma^+ \rightarrow \Gamma$ is the
unique extension.

In general, a renaming $I_{\pi} : \pi \cdot \Gamma \rightarrow \Gamma$ will
not be an extension unless $\pi$ is the identity on the labels from $\Gamma$.
In our work, both renaming and extentions are used together, so we want to
consider a broader class of nominal assignments that include both extensions
and renamings.

\begin{definition}
  Those nominal assignments where every label occurs at most once will
  be called general extensions. 
\end{definition}

We see that for every permutation of labels $\pi :
\V \rightarrow \V$ and every context $\Gamma$ that $I_{\pi}
\left( \Gamma \right) : \pi \cdot \Gamma \rightarrow \Gamma$ is a general
extension (and hence also a nominal assignment).

\begin{theorem}
  \label{genExtPlusRename}Every general extension $A : \Gamma^+ \rightarrow
  \Delta$ can be turned into an extension by composing it with an appropriate
  renaming.
\end{theorem}

The proof of this theorem, along with all other theorems, lemmas and 
corollaries in this section can be found in Appendix~\ref{AppA}.

\begin{corollary}\label{cor:ExtDecomp}
  Every general extension $A : \Gamma^+ \rightarrow \Delta$ can be decomposed
  into an extension $A_e : \Gamma^+ \rightarrow \Gamma$ followed by a renaming
  $A_r : \Gamma \rightarrow \Delta$.
\end{corollary}

\noindent These general extensions form a category which plays an important
r\^ole.

\begin{minipage}{0.3\textwidth}
\begin{tikzpicture}[node distance=2cm, auto]
  \node (P) {$\Gamma^{+}$};
  \node (B) [right of=P] {$\Delta^{+}$};
  \node (A) [below of=P] {$\Gamma$};
  \node (C) [below of=B] {$\Delta$};
  \draw[->] (P) to node {$f^{+}$} (B);
  \draw[->] (P) to node [swap] {$A$} (A);
  \draw[->] (A) to node [swap] {$f^{-}$} (C);
  \draw[->] (B) to node {$B$} (C);
\end{tikzpicture}
\end{minipage}
\begin{minipage}{0.69\textwidth}
\begin{definition}
  The category of general extensions $\E$ has all general extensions
  from $\B$ as objects, and given two general extensions $A :
  \Gamma^+ \rightarrow \Gamma$ and $B : \Delta^+ \rightarrow \Delta$, an arrow
  $f : A \rightarrow B$ is a commutative square from $\B$.
  We will denote this commutative square by $\left\langle f^+, f^-
  \right\rangle : A \rightarrow B$.
\end{definition}
\end{minipage}

\noindent We remind the reader of the usual convention in category theory
where arrows include their domain and codomain as part of their structure
(which we implicitly use in the definition above).

\begin{lemma}\label{lem:InitialIso}
  Every general extension is isomorphic in $\E$ to
  an extension $B : \Gamma^{\circ} \rightarrow \Gamma$ where $\Gamma$ is an
  initial segment of $\Gamma^{\circ}$.
\end{lemma}

This category of general extensions $\E$ is fibered over the category
$\B$ by the codomain functor $\tmop{cod} : \E \rightarrow
\B$. \ Given general extensions $A : \Gamma^+ \rightarrow \Gamma^{}$
and $B : \Delta^+ \rightarrow \Delta$ and a morphism $\left\langle f^+, f^-
\right\rangle : A \rightarrow B$ in $\E$ we have
\[
  \tmop{cod} \left( A \right) \assign \Gamma\qquad \qquad
  \tmop{cod} \left( f \right) \assign f^-
\]
\begin{theorem}
  \label{codIsFibration}The functor $\tmop{cod} : \E \rightarrow
  \B$ is a fibration.
\end{theorem}

\begin{corollary}\label{genExtFromCartLift}
  Given $u : \Gamma \rightarrow \Delta$, a general extension $A : \Delta^+
  \rightarrow \Delta$, and a cartesian lifting $\bar{u} \left( A \right) :
  u^{\ast} \left( A \right) \rightarrow A$, if $u$ is a general extension, then
  $\bar{u} \left( A \right)^+$ is also a general extension.
\end{corollary}

\begin{example}
  The nominal assignment (and general extension)
  \[ u \assign \left[ \begin{array}{rcl}
       U & \mapsto & U\\
       \left( * \right) & \mapsto & \left( + \right)\\
       \tmop{associative} & \mapsto & \tmop{associative}
     \end{array} \right] : \text{\tmtexttt{AbelianSemigroup}} \rightarrow
     \text{\tmtexttt{Semigroup}} \]
  and the extension $A
  \assign \delta_{\text{\tmtexttt{Semigroup}}} : \text{\tmtexttt{Monoid}}
  \rightarrow \text{\tmtexttt{Semigroup}}$ induce the existence
  (via theorem~\ref{codIsFibration}) of
  some Cartesian lifting $\bar{u} \left( A \right) : u^{\ast}
  \left( A \right) \rightarrow A$ in $\E$. One example of such a
  Cartesian lifting for $\bar{u}$ is
  
\begin{center}
\begingroup
\everymath{\scriptstyle}
\begin{tikzpicture}[node distance=3.6cm, auto]
  \node (CM) {\tt{AbelianMonoid}};
  \node (M) [right of=CM] {\tt{Monoid}};
  \node (CS) [below of=CM, node distance=1.8cm] {\tt{AbelianSemigroup}};
  \node (S) [below of=M, node distance=1.8cm] {\tt{Semigroup}};
  \draw[->] (CM) to node [swap] {$\bar{u}(A)^{+}$} (M);
  \draw[->] (CM) to node [swap] {$u^{*}(A)$} (CS);
  \draw[->] (M) to node {$A$} (S);
  \draw[->] (CS) to node {$u$} (S);
\end{tikzpicture}
\endgroup
\end{center}
  
  \noindent where $\texttt{AbelianMonoid}$ is
  \[ \left\langle
     \begin{array}{rcl}
       U & : & \tmop{Type}\\
       0 & : & U\\
       \left( + \right) & : & U \times U \rightarrow U\\
       \tmop{rightIdentity} & : & \forall x : U.x + 0 = x\\
       \tmop{leftIdentity} & : & \forall x : U. 0 + x = x\\
       \tmop{associative} & : & \forall x, y, z : U. \left( x + y \right) + z
       = x + \left( y + z \right)\\
       \tmop{commutative} & : & \forall x, y : U. \left( x + y \right) =
       \left( y + x \right)
     \end{array} \right\rangle \]
  and $u^{\ast} \left( A \right) : \text{\tmtexttt{AbelianMonoid}}
  \rightarrow \text{\tmtexttt{AbelianSemigroup}}$ is the diagonal
  assignment, where
    $\bar{u} \left( A \right)^+ : \text{\tmtexttt{AbelianMonoid}}
    \rightarrow \text{\tmtexttt{Monoid}}$ is
  \begin{eqnarray*}
    \bar{u} \left( A \right)^+ & \assign & \left[ \begin{array}{rcl}
      U & \mapsto & U\\
      e & \mapsto & 0\\
      \left( * \right) & \mapsto & \left( + \right)\\
      \tmop{rightIdentity} & \mapsto & \tmop{rightIdentity}\\
      \tmop{leftIdentity} & \mapsto & \tmop{leftIdentity}\\
      \tmop{associative} & \mapsto & \tmop{associative}
    \end{array} \right]
  \end{eqnarray*}
\end{example}

\begin{wrapfigure}[12]{l}{0pt}
\begin{tikzpicture}[node distance=3.2cm, auto]
  \node (UV) {\small $\left\langle\begin{array}{rcl}
               U & : & Type \\
               U' & : & Type 
               \end{array}\right\rangle$};
  \node (U1) [right of=UV] {\small $\langle U : Type \rangle$};
  \node (U2) [below of=UV, node distance=2.2cm] {\small $\langle U : Type \rangle$};
  \node (T) [below of=U1, node distance=2.2cm] {\small $\langle \rangle$};
  \node (U3) [node distance=1.6cm, left of=UV, above of=UV] {\small $\langle U : Type \rangle$};
  \draw[->] (UV) to node {$\bar{u}(u)^{+}$} (U1);
  \draw[->] (UV) to node {$u^{*}(u)$} (U2);
  \draw[->] (U2) to node [swap] {$u$} (T);
  \draw[->] (U1) to node {$u$} (T);
  \draw[->, bend right] (U3) to node [swap] {$id$} (U2);
  \draw[->, bend left] (U3) to node {$id$} (U1);
  \draw[->, dashed] (U3) to node {$f$} (UV);
\end{tikzpicture}
\end{wrapfigure}
In almost all of the development of the algebraic hierarchy, the nominal
assignments that we use are all general extensions. However, it is important
to note that the definition of a Cartesian lifting requires nominal
assignments that are not necessarily general extensions, even if all the
inputs are general extensions.

Consider the simple case (pictured above) where $u : \left\langle U :
\tmop{Type} \right\rangle \rightarrow \left\langle \right\rangle$ is the unique
extension, and a Cartesian lifting of $u$ over itself. The mediating arrow for
$\tmop{id} : \left\langle U : \tmop{Type} \right\rangle \rightarrow
\left\langle U : \tmop{Type} \right\rangle$ and itself must be
\begin{eqnarray*}
  f & : & \left\langle U : \tmop{Type} \right\rangle \rightarrow \left\langle
  U : \tmop{Type} ; U' : \tmop{Type} \right\rangle\\
  f & \assign & \left[ U \mapsto U, U' \mapsto U \right]
\end{eqnarray*}
which is not a general extension.

\section{Semantics of Theory Presentation Combinators}\label{sec:tpc2}

Like in the previous section, we will assume that we have a background
type theory with well-formedness \emph{judgments}, which defines four different
sorts, namely 
$\left(\mathsf{Type}, \mathsf{Term}, \mathsf{Kind}, \mathsf{Prop}\right)$.
The symbols used in the type theory itself will be called \emph{labels},
whereas the symbols used for theory presentations will be called \emph{names}.
As above, $a \mapsto b$ denotes a \emph{substitution}.  Using this, we
can define the formal syntax for our combinators as follows.

\begin{minipage}{.42\textwidth}
\begin{align*}
a,b,c &\in \text{labels} & \tau &\in \mathsf{Type} \\
A,B,C &\in \text{names} & k &\in \mathsf{Kind} \\
l &\in \text{judgments}^{*} & t &\in \mathsf{Term}  \\
r &\in \left(a_i\mapsto b_i\right)^{*} & \theta &\in \mathsf{Prop} \\
\end{align*}
\end{minipage}
\begin{minipage}{.57\textwidth}
\begin{align*}
\mathsf{tpc} \Coloneqq &\  \mathsf{extend}\  A\  \mathsf{by}\  \{ l \} \\
 | &\  \mathsf{combine}\  A\  r_1,\  B\  r_2 \\
 | &\  A\ ;\ B \\
 | &\  A\ r \\
 | &\  \mathsf{Empty} \\
 | &\  \mathsf{Theory\ }\{ l \}
\end{align*}
\end{minipage}

Intuitively, the six forms correspond to: 
extending a theory with new knowledge, combining two theories into a 
larger one, sequential composition of theories, renaming, a constant
for the Empty theory, and an explicit theory.

What we do next is slightly unusual: rather than give a single denotational
semantics, we will give \emph{two}, one in terms of objects of 
$\B$, and one in terms of objects of $\E$ (which are special
arrows in $\B$).  In fact, we have a third semantics, in terms of 
(partial) Functors over the contextual category, but we will omit it for
lack of space.
First, we give the semantics in terms of objects of $\B$, where 
$\semP{-}$ is the (obvious) semantics in $\V\rightarrow\V$ of a
renaming.\\
\begin{minipage}{.42\textwidth}
\begin{align*}
\semC{-} :\ & \mathsf{tpc} \partialf |\B| \\
\semC{\mathsf{Empty}} =\  & \EmpThy\\
\semC{\mathsf{Theory\ }\{ l \}} \cong\  & \langle l \rangle \\
\semC{A\ r} =\  & \semP{r} \cdot \semC{A} \\
\semC{A ; B} =\  & \semC{B} \\
\semC{\mathsf{extend\ }A\mathsf{\ by\ }\{ l \}} \cong\  &
    \semC{A} \ctxcat \ \langle l\rangle \\
\semC{\mathsf{combine\ } A_1 r_{1}, A_2 r_{2}} \cong\  &
    D
\end{align*}
\end{minipage}
\begin{minipage}{.49\textwidth}
\begin{center}
\begin{tikzpicture}[node distance=2.4cm, auto]
  \node (D) {$D$};
  \node (A1) [right of=D] {$A_1$};
  \node (A2) [below of=D] {$A_2$};
  \node (A) [below of=A1] {$A$};
  \draw[->] (D) to node {$\semP{r_1}\circ\delta_{A_1}$} (A1);
  \draw[->] (D) to node {$\semP{r_2}\circ\delta_{A_2}$} (A2);
  \draw[->] (A2) to node [swap] {$\delta_{A}$} (A);
  \draw[->] (A1) to node {$\delta_{A}$} (A);
\end{tikzpicture}
\end{center}
\end{minipage}
\noindent where $D$ comes from the (potential) pullback diagram on the
right, in which we omit $\semC{-}$ around the $A$s for clarity.  We use $\cong$
to abbreviate ``when the rhs is a well-formed context''.
For the semantics of $\mathsf{combine}$, it must be the case where the diagram
at right is a \emph{pullback} (in $\B$), where $A$ is the greatest 
lower bound context $\semC{A_1}\sqcap\semC{A_2}$.
Furthermore $\semP{r_1}$ and $\semP{r_2}$ must leave $A$ 
invariant.  We remind the reader of the requirement for these renamings:
the users must pick which cartesian lifting they want, and this cannot
be done automatically (as demonstrated at the end of last section).

The second semantics, is in terms of the
objects of $\E$, in other words, the \emph{special} arrows of $\B$, 
as defined in Section~\ref{sec:category}.

\begin{minipage}{.60\textwidth}
\begin{align*}
\semE{-} & :\ \mathsf{tpc} \partialf |\E| \\
\semE{\mathsf{Empty}} & =\ \mathsf{id}_{\EmpThy} \\
\semE{\mathsf{Theory\ }\{ l \}} & \cong\  !_{\langle l \rangle} \\
\semE{A\ r} & =\ \semP{r} \cdot \semE{A} \\
\semE{A ; B} & =\  \semE{A} \circ \semE{B} \\
\semE{\mathsf{extend\ }A\mathsf{\ by\ }\{ l \}} & \cong\  \delta_{A} \\
\semE{\mathsf{combine\ } A_1 r_{1}, A_2 r_{2}} & \cong\  
  \semP{r_1}\circ \delta_{T_1}\circ \semE{A_1} \\
  & \cong\ \semP{r_2}\circ \delta_{T_2}\circ \semE{A_2} \\
\end{align*}
\end{minipage}
\begin{minipage}{.39\textwidth}
\begin{center}
\begin{tikzpicture}[node distance=2.4cm, auto]
  \node (D) {$D$};
  \node (A1) [right of=D] {$T_1$};
  \node (A2) [below of=D] {$T_2$};
  \node (A) [below of=A1] {$T$};
  \draw[->] (D) to node {$\semP{r_1}\circ \delta_{T_1}$} (A1);
  \draw[->] (D) to node {$\semP{r_2}\circ \delta_{T_2}$} (A2);
  \draw[->] (A2) to node [swap] {$A_2$} (A);
  \draw[->] (A1) to node {$A_1$} (A);
\end{tikzpicture}
\end{center}
\end{minipage}
\noindent 
The diagram on the right has to be verified to be a pullback diagram
(this is why the semantics is partial here too).
Here we assume $\semE{A_1} \in Hom(T_1,T)$ and
$\semE{A_2}\in Hom(T_2,T)$, and that both $\semP{r_1}$ and $\semP{r_2}$
leave $T$ invariant.

\begin{theorem}
For all $\mathsf{tpc}$ terms except $\mathsf{combine}$, 
$\semC{s} = \mathsf{dom} \semE{s}$.  When
$s \equiv \mathsf{combine\ } A_{1} r_{1}, A_{2} r_{2}$, if 
$\mathsf{cod}\left(\semE{A_1}\right) = 
\mathsf{cod}\left(\semE{A_2}\right) = \semC{A_1}\sqcap\semC{A_2}$,
and neither arrows $\semE{A_1}$ nor $\semE{A_2}$ involve renamings,
then $\semC{s} = \mathsf{dom} \semE{s}$ in that case as well.
\end{theorem}

\noindent The proof is a straightforward comparison of the semantic
equations.  This theorem basically says that, as long as we only use
$\mathsf{combine}$ on the ``natural'' base of two arrows which are pure
extensions, our semantics are compatible. In a \emph{tiny theories} setting,
this can be arranged.

\section{Discussion}\label{sec:discussion}

It is important to note that we are essentially parametric in the 
underlying type theory.  This should allow us to be able to generalize our
work in ways similar to Kohlhase and Rabe's MMT~\cite{MMT}.

The careful reader might have notice that in the syntax of 
section~\ref{sec:tpc1}, our \thy{combine} had an \thy{over} keyword.  This
allowed our previous implementation~\cite{MathSchemeExper} to come 
partway to the $\E$ semantics above.  This is a straightforward extension
to the semantics: 
$\semC{\mathsf{combine\ } A_1 r_{1}, A_2 r_{2} \ \mathsf{over\ }C}$
would replace $A = \semC{A_{1}}\sqcap\semC{A_{2}}$ with $\semC{C}$,
with corresponding adjustments to the rest of the pushout diagram.
For $\semE{-}$, one would insist that
$\mathsf{cod}\left(\semE{A_1}\right) = 
\mathsf{cod}\left(\semE{A_2}\right) = \semC{C}$.

What is more promising\footnote{Work in progress} still is that most of our
terms can also be interpreted as \emph{Functors} between fibered categories.
This gives us a semantic for each term as a ``construction'', which can be
reused (as in our example with commutativity in section~\ref{sec:tpc1}).
Furthermore, since fibered categories interact well will limits and
colimits, we should also be able to combine constructions and diagrams
so as to fruitfully capture further structure in theory hierarchies.

It should also be noted that our work also extends without difficulty to
having \emph{definitions} (and other such conservative extensions) in
our contexts.  This is especially useful when transporting theorems from
one setting to another, as is done when using the 
``Little Theories'' method~\cite{LittleTheories}.
We also expect our work to extend to allow Cartesian liftings of
extensions over arbitrary assignments (aka views) from the full category of
contexts.

Lastly, we have implemented a ``flattener'' for our semantics, which
basically turns a presentation $A$ into a flat presentation 
$\mathsf{Theory}\{\ l\}$ by computing $\mathsf{cod}\left(\semE{A}\right)$.
This fulfils our second requirement, where the method of construction
of a theory is invisible to users of flat theories.

\section{Related Work}\label{sec:related}

We will not consider work in universal algebra, institutions or 
categorical logic as ``related'', since they employ
$\alpha$-equivalence on labels (as well as on bound variables), which we
consider un-helpful for theory presentations meant for human consumption.  We
also leave aside much interesting work on dependent record types (which we
use), as these are but one implementation method for theories, and we consider
\emph{contexts} as a much more fundamental object.

We have been highly influenced by the early work of Burstall and 
Goguen~\cite{PuttingTheoriesTogether,SemanticsOfClear}, and Doug Smith's
Specware~\cite{Smith93,Smith99}.  They gave us the basic semantic tools
we needed.  But we quickly found out, much to our dismay, that neither of
these approaches seemed to scale very well.  Later, we were hopeful that 
CASL~\cite{CoFI:2004:CASL-RM} might work for us, but then found that
their own base library was improperly factored and full of redundancies.
Of the vast algebraic specification literature around this topic, we want
to single out the work of Oriat~\cite{Oriat} on isomorphism of 
specification graphs as capturing similar ideas to ours on extreme
modularity.  And it cannot be emphasized enough how
crucial Bart Jacob's book~\cite{JacobsCLTT} has been to our work.

From the mathematical knowledge management side, it should be clear
that MMT~\cite{MMT} is closely related.  The main differences are that they
are quite explicit about being foundations-independent (it is implicit
in our work), they use long names, and their theory operations are mostly
theory-internal, while ours are external.  This makes a big difference,
as it allows us to have multiple semantics, while theirs has to be fixed.
And, of course, the work presented in the current paper covers just a small
part of the vast scope of MMT.

There are many published techniques and implementations of algebraic
hierarchies in dependently typed proof assistants including
\cite{GeuversWiedijk,SpittersvanderWeegen,GarillotGonthier,CoenTassi2009}.
Our work does not compete with these implementations, but rather complements
them.  More specifically, we envision our work as a meta-language which can
be used to specify algebraic hierarchies, which can subsequently be implemented
by using any of the aforementioned techniques.  In
particular we note that maintaining the correct structures for packed-classes
of \cite{GarillotGonthier} is particularly difficult, and deriving the required
structures from a hierarchy specification would alleviate much of this
burden.  Other cited work, (for example~\cite{CoenTassi2009}) focus on other 
difficult problems such as \emph{usability}, via providing coercions and
unification hints to match particular terms to theories. Even though some
similar techniques (categorical pullbacks) are used in a similar context,
the details are very different.

\section{Conclusion}\label{sec:conclusion}
There has been a lot of work done in mathematics to give structure to
mathematical theories, first via universal algebra, then via category
theory (e.g. Lawvere theories).  But even though a lot of this work 
started out being somewhat syntactic, very quickly it became mostly
semantic, and thus largely useless for the purposes of concrete
implementations.

We make the observation that, with a rich enough type theory, we
can identify the category of theory presentations with the opposite of the
category of contexts.  This allows us to draw freely from developments in
categorical logic, as well as to continue to be inspired by algebraic
specifications.  Interestingly, key here is to make the opposite choice as
Goguen's in two ways: our base language is firmly higher-order, while our
``module'' language is first-order, and we work in the opposite category.

We provide a simple-to-understand term language of ``theory expression
combinators'', along with multiple (categorical) semantics.  We have shown
that these fit our requirements of allowing to capture mathematical
structure, while also allowing this structure to be hidden from users.

Even more promising, our use of very standard categorical constructions 
points the way to simple generalizations which should allow us to capture
even more structure, without having to rewrite our library.  Furthermore,
as we are independent of the details of the type theory, this structure
seems very robust, and our combinators should thus port easily to other
systems.

\bibliography{mathscheme} 
\bibliographystyle{splncs03}
 
\appendix

\section{Proofs}\label{AppA}

Given a nominal assignment of the form 
$\ren{y_i}{x_{a\left(i\right)}}{i=0}{m-1}$,
we call the function $a : \N_{\left| \Delta \right|}
\rightarrow \N_{\left| \Gamma \right|}$ above an {\tmdfn{indexing
function}}. The indexing function of a nominal assignment is sufficient to
specify the nominal assignment. Given a morphism $A : \Gamma \rightarrow
\Delta$ with indexing function $a : \N_{\left| \Delta \right|}
\rightarrow \N_{\left| \Gamma \right|}$ and a morphism $B : \Delta
\rightarrow \Xi$ with indexing function $b : \N_{\left| \Xi \right|}
\rightarrow \N_{\left| \Delta \right|}$, that the composition $B
\circ A : \Gamma \rightarrow \Xi$ has $a \circ b : \N_{\left| \Xi
\right|} \rightarrow \N_{\left| \Gamma \right|}$ for its indexing
function.

\begin{proof}[of Theorem~\ref{genExtPlusRename}, Section~\ref{sec:category}]
  Suppose $A = \ren{y_i}{x_{a\left(i\right)}}{i=0}{m-1}$. Because $a$ is 
  injective for general extensions, we
  can select a permutation of names $\pi_A : \V \rightarrow
  \V$ such that
  $\pi_A \left( y_i \right) = x_{a \left( i \right)}$.
  This implies that $\pi_A \cdot \Delta$ is a sub-context of $\Gamma^+$.
  Consider the renaming $I_{\pi_A^{- 1}} \left( \pi_A \cdot \Delta \right)
  : \Delta \rightarrow \pi_A \cdot \Delta$. We see that the composition
  $I_{\pi_A^{- 1}} \left( \pi_A \cdot \Delta \right) \circ A : \Gamma^+
  \rightarrow \pi_A \cdot \Delta$ is an extension since
  \begin{eqnarray*}
    I_{\pi_A^{- 1}} \left( \pi_A \cdot \Delta \right) \circ A & = & 
      \ren{\pi_A\left(y_i\right)}{y_i}{i=0}{m-1} \circ \ren{y_i}{x_{a\left(i\right)}}{i=0}{m-1} \\
    & = & \ren{\pi_A\left(y_i\right)}{x_{a\left(i\right)}}{i=0}{m-1} = 
          \ren{x_{a\left(i\right)}}{x_{a\left(i\right)}}{i=0}{m-1}
  \end{eqnarray*}
  is the diagonal assignment. \qed
\end{proof}

\begin{proof}[of Corollary~\ref{cor:ExtDecomp}, Section~\ref{sec:category}]
  From the previous theorem we know that there is a permutation $\pi_A :
  \V \rightarrow \V$ such that $I_{\pi_A^{- 1}} \left( \pi_A
  \cdot \Delta \right) \circ A : \Gamma^+_{} \rightarrow \pi_A \cdot
  \Delta$ is an extension. Let $\Gamma \assign \pi_A \cdot \Delta$ and let
  $A_e \assign I_{\pi_A^{- 1}} \left( \pi_A \cdot \Delta \right) \circ A$.
  Then we can take $A_r \assign I_{\pi_A^{}} \left( \Delta \right) : \Gamma
  \rightarrow \Delta$. We immediately see that
  \begin{eqnarray*}
    A_r \circ A_e & = & I_{\pi_A} \left( \Delta \right) \circ I_{\pi_A^{-
    1}} \left( \pi_A \cdot \Delta \right) \circ A\\
    & = & I_{\pi_A} \left( \Delta \right) \circ \left( I_{\pi_A} \left(
    \Delta \right) \right)^{- 1} \circ A = A.
  \end{eqnarray*}
\qed
\end{proof}

\begin{proof}[of Lemma~\ref{lem:InitialIso}, Section~\ref{sec:category}]
  By Theorem~\ref{genExtPlusRename} every general extension is isomorphic to
  an extension, so it suffices to show that every extension $A : \Gamma^+
  \rightarrow \Gamma$ is isomorphic to an extension $B : \Gamma^{\circ}
  \rightarrow \Gamma$ where $\Gamma$ is an initial segment of
  $\Gamma^{\circ}$. If $\Gamma$ is already an initial segment of $\Gamma^+$,
  then we can just take $\Gamma^{\circ} \assign \Gamma^+$. Otherwise there
  exist $i$ and $j$ such that $i < j < s$ and
  \[ \Gamma^+ = \left\langle 
     x_0 : \tau_0 ; \ldots ; 
     x_i : \tau_i ; \ldots ;
     x_j : \tau_j ; \ldots ; 
     x_{s - 1} : \tau_{s - 1} \right\rangle \]
  where $x_{i - 1}$ is not in $\Gamma$, $x_i, \ldots, x_{j - 1}$ are all in
  $\Gamma$, and $x_j, \ldots, x_{s - 1}$ are all not in $\Gamma$. Because
  $\Gamma$ is a well formed context, it must be the case that $x_{i - 1}$ does
  not occur in $\tau_i, \ldots, \tau_{j - 1}$, and we can safely rearrange
  $\Gamma^+$ into
  \begin{multline*}
  \hat{\Gamma}^{+}= \langle x_0 : \tau_0 ; \ldots ; x_{i - 2} : \tau_{i -
     2}, x_i : \tau_i ; \ldots ; x_{j - 1} : \tau_{j - 1}, \\
     x_{i - 1} : \tau_{i - 1}, x_j : \tau_j ; \ldots ; 
     x_{s - 1} : \tau_{s - 1} \rangle .
  \end{multline*}
  This new context $\hat{\Gamma}^{+}$ is isomorphic to $\Gamma^+$ via
  $\delta_{\Gamma^+} : \hat{\Gamma}^{+} \rightarrow \Gamma^+$. Continuing by
  induction on $i$, we can eventually construct a $\Gamma^{\circ}$ which is
  isomorphic to $\Gamma^+$ via $\delta_{\Gamma^+} : \Gamma^{\circ} \rightarrow
  \Gamma^+$ such that $\Gamma$ is an initial segment of $\Gamma^{\circ}$. \
  Therefore $A : \Gamma^+ \rightarrow \Gamma$ and $A \circ \delta_{\Gamma^+} :
  \Gamma^{\circ} \rightarrow \Gamma$ are isomorphic in $\E$. \qed
\end{proof}

\noindent\textit{Proof. (of Theorem~\ref{codIsFibration}, 
Section~\ref{sec:category})}
  To prove that $\tmop{cod} : \E \rightarrow \B$ is a
  fibration we need to show that for any nominal assignment $u : \Gamma
  \rightarrow \Delta$ from $\B$ and a general extension $A : \Delta^+
  \rightarrow \Delta$, there exists a Cartesian lifting $\bar{u} \left( A
  \right) : u^{\ast} \left( A \right) \rightarrow A$ in $\E$.
  
  We need to show that $\bar{u} \left( A \right)$ is a pullback diagram of the
  cospan\\ $\Gamma \overset{u}{\longrightarrow} \Delta
  \overset{A}{\longleftarrow} \Delta^+$ in $\B$ and that $u^{\ast}
  \left( A \right)$ is a general extension. By Lemma~\ref{lem:InitialIso} we can
  assume that $A$ is an extension and $\Delta$ is an initial segment of
  $\Delta^+$. We can further suppose that the names in
  $\Delta^+$ are disjoint from the names $\Gamma$,
  by applying a suitable permutation $\pi$
  to $A$ and noting that if $\Delta$ is an initial segment of
  $\Delta^+$ then $\pi \cdot \Delta$ is an initial segment of $\pi \cdot
  \Delta^+$.
  
  Suppose $\Delta^+ = \ctx{y}{\sigma}{0}{m+e-1}$ and
  $u \assign \ren{y_i}{x_{a\left(i\right)}}{i=0}{m-1}$.
  Define an extension of $\Gamma$,
  $\Gamma^+ \assign \ctx{x}{\tau}{0}{n+e=1}$, where
    $x_{n + k} \assign y_{m + k}$ and
    $ \tau_{n + k} \assign \sigma_{m + k} \ren{y_i}{x_{a\left(i\right)}}{i=0}{m-1}$
  for all $k < e$. Let $u^{\ast} \left( A \right) \assign \delta_{\Gamma} :
  \Gamma^+ \rightarrow \Gamma$. By definition $u^{\ast} \left( A \right)$ is
  an extension (and $\Gamma$ is an initial segment of $\Gamma^+$). \ We can
  define a nominal assignment $u^+ \left( A \right) : \Gamma^+ \rightarrow
  \Delta^+$ by $u^+ \left( A \right) \assign 
    \ren{y_i}{x_{a\left(i\right)}}{i=0}{m-1} \rencat \ren{y_{m+i}}{x_{n+i}}{i=0}{e-1}$.
  By construction, $u \circ u^{\ast} \left( A \right) = A \circ u^+ \left( A
  \right)$, and so $\bar{u} \left( A \right) \assign \left\langle u^+, u
  \right\rangle : u^{\ast} \left( A \right) \rightarrow A^{}$ is a commutative
  square.
  
  To show that $\bar{u} \left( A \right)$ is a Cartesian lifting of $u$ over
  $A$, consider some $X : \Xi^+ \rightarrow \Xi$ in $\E$ along with
  $g : X \rightarrow A$ and $v : \Xi \rightarrow \Gamma$ such that $u \circ v
  = \tmop{cod} \left( g \right)$. \ Again, without loss of generality, we can
  assume that $X$ is an extension and $\Xi$ is an initial segment of $\Xi^+$.
  
  Suppose
  $\Xi = \ctx{z}{\xi}{0}{r-1}, \Xi^+ = \ctx{z}{\xi}{0}{r+s-1},
   v = \ren{x_j}{z_{b\left(j\right)}}{j=0}{n-1}, g^{+} = \ren{y_i}{z_{c\left(i\right)}}{i=0}{m+e-1}$.
  We can define $v^+ : \Xi^+ \rightarrow \Gamma^+$ as
  $v \rencat \ren{x_j}{z_{c\left(j\right)}}{j=m}{m+e-1}.$
  To see that $v^+$ is well formed, recall that
  $\tau_{n + k} = \sigma_{n + k} \ren{y_i}{x_{a\left(i\right)}}{i=0}{m-1}.$
  Because for all $i < m$, $c \left( i \right) = b \left( a \left( i \right)
  \right)$,
  \[ \tau_{n+k}\left[v^+\right] =\ \tau_{n+k}\ 
       \left[v \rencat \ren{x_j}{z_{c\left(j\right)}}{j=m}{m+e-1}\right]
     =\ \sigma_{m+k} \ren{y_i}{z_{c\left(i\right)}}{i=0}{m+k-1} \]
  
\begin{wrapfigure}[11]{l}{0pt}
\begin{tikzpicture}[node distance=2cm, auto]
  \node (P) {$\Gamma^{+}$};
  \node (B) [right of=P] {$\Delta^{+}$};
  \node (A) [below of=P] {$\Gamma$};
  \node (C) [below of=B] {$\Delta$};
  \node (X) [node distance=1.4cm, left of=P, above of=P] {$\Xi^{+}$};
  \node (Y) [below of=X] {$\Xi$};
  \draw[->] (P) to node {$u^{+}$} (B);
  \draw[->] (P) to node {$u^{*}(A)$} (A);
  \draw[->] (A) to node [swap] {$u$} (C);
  \draw[->] (B) to node {$A$} (C);
  \draw[->] (X) to node [swap] {$X$} (Y);
  \draw[->, bend left] (X) to node {$g^{+}$} (B);
  \draw[->, dashed] (X) to node {$v^{+}$} (P);
  \draw[->] (Y) to node [swap] {$v$} (A);
\end{tikzpicture}
\end{wrapfigure}
\noindent  which is well typed in $\Xi$ because $g^+$ is well-defined.
  Clearly, $u^{\ast} \left( A \right) \circ v^+ = v \circ X$, and $g^+ = u^+
  \left( A \right) \circ v^+$ by construction. Therefore, $\bar{v} \assign
  \left\langle v^+, v \right\rangle : X \rightarrow u^{\ast} \left( A \right)$
  is an arrow in $\E$ such that $\bar{u} \circ \bar{v} = g$.

  Finally, we need to show that $\bar{v}$ is the unique arrow in $\E$
  such that $\bar{u} \circ \bar{v} = g$ and $\tmop{cod} \left( \bar{v} \right)
  = v$. \ Suppose $\bar{w} : X \rightarrow u^{\ast} \left( A \right)$ is
  another arrow such that $\bar{u} \circ \bar{w} = g$ and $\tmop{cod} \left(
  \bar{w} \right) = v$. Say
  $\bar{w}^+ \assign \ren{x_j}{z_{d\left(j\right)}}{j=0}{n+e-1}.$
  Because $u^{\ast} \left( A \right) \circ \bar{w}^+ = X \circ v$, it must be
  the case that for all $i < n$, $d \left( i \right) = b \left( i \right)$.
  Because $u^+ \left( A \right) \circ \bar{w}^+ = g^+$, it must be the case
  that for all $k < e$, $d \left( n + k \right) = c \left( m + k \right)$.
  Therefore, $\bar{w}^+ = \bar{v}^+$, and so $\bar{w} = \bar{v}$.
\qed

\begin{proof}[of Corollary~\ref{genExtFromCartLift}, 
Section~\ref{sec:category}]
  For such a $u$, a cartesian lifting $\bar{A} \left( u
  \right) : A^{\ast} \left( u \right) \rightarrow u$ is isomorphic to the
  transpose of $\bar{u} \left( A \right)$, and in particular $\bar{u} \left( A
  \right)^+$ will be isomorphic to $A^{\ast} \left( u \right)$. \ Since
  $A^{\ast} \left( u \right)$ is a general extension, then so is $\bar{u}
  \left( A \right)^+$. \qed
\end{proof}

\end{document}